\documentclass[10pt]{article}

\usepackage{amsmath}
\usepackage{amssymb}

\usepackage{graphicx}

\usepackage{cite}

\usepackage{color} 

\usepackage{amsfonts}

\usepackage[labelfont=bf,labelsep=period,justification=raggedright]{caption}

\makeatletter
\renewcommand{\@biblabel}[1]{\quad#1.}
\makeatother

\date{}

\newcommand{\dom}{{\cal D}}
\newcommand{\SI}{materials and methods}
\newcommand{\balanceeq}{\ref{eq:balance}}

\newcommand{\mk}{\langle k\rangle}

\newcommand{\figtimeseries}{\ref{figure_main1}}
\newcommand{\eqmatrix}{8}

\begin{document}

\begin{flushleft}
{\Large
\textbf{A self-organized model for cell-differentiation based on variations of molecular decay rates}
}
\\
Rudolf Hanel$^{1}$, 
Manfred P\"ochacker$^{1}$, 
Manuel Sch\"olling$^{1}$, 
Stefan Thurner$^{1,2,\ast}$
\\
\bf{1} Section for Science of Complex Systems/CeMSIIS, Medical University of Vienna, Spitalgasse 23, A-1090, Vienna, Austria, 
\\
\bf{2} Santa Fe Institute, 1399 Hyde Park Road, Santa Fe, NM 87501; USA
\\
$\ast$ E-mail: stefan.thurner@meduniwien.ac.at
\end{flushleft}

\section*{Abstract}
Systemic properties of living cells are the result of molecular dynamics  governed by so-called genetic regulatory networks (GRN). 
These networks capture all possible features of cells and are responsible for the immense levels of adaptation characteristic to living systems. At any point in time only small subsets of these networks are active. 
Any active subset of the GRN leads to the expression of particular sets of  molecules (expression modes).  
The subsets of active networks change over time, leading to the observed complex dynamics of expression patterns. 
Understanding of this dynamics becomes increasingly important in systems biology and medicine. 
While the importance of transcription rates and catalytic interactions has been widely recognized in modeling 
genetic regulatory systems, the understanding of the role of degradation of biochemical agents (mRNA, protein) 
in regulatory dynamics remains limited. 
Recent experimental data suggests that there exists a functional relation between 
mRNA and protein decay rates and expression modes. 
In this paper we propose a model for the dynamics of successions of sequences of active subnetworks of the GRN. The  
model is able to reproduce key  characteristics of molecular dynamics, including homeostasis, multi-stability, periodic dynamics, 
alternating activity, differentiability, and  self-organized critical dynamics.
Moreover the model allows to naturally understand the mechanism behind the relation between 
decay rates and expression modes. 
The model explains recent experimental observations that decay-rates (or turnovers) vary between differentiated tissue-classes
at a general systemic level and highlights the role of intracellular  decay rate control mechanisms in cell differentiation.


\section*{Introduction}
Understanding living cells at a systemic level is an increasingly important 
challenge in biology and medicine \cite{Hood2004,Gavin2006,Kashtan2005,Chassey2008,Church2005}. 
Regulatory interactions between intracellular molecular agents 
(e.g. DNA, RNA, proteins, hormones, trace elements), form so-called {\it genetic regulatory networks} (GRN), 
which orchestrate gene expression and replication, coordinate metabolic activity, 
and cellular development, respond to changes in the environment, or stress. 
GRN coordinate regulatory dynamics on all levels from
cell-fate \cite{Greer2005, Tothova2007} to stress response \cite{Burg1996,Capaldi2008,Pirkkala2001}.
Qualitative understanding of GRN topology is for instance obtained from promoter 
sequences \cite{Beer2004,Banerjee2004,Gerz2009},
gene-expression profiling \cite{Vokes2008,Visel2009,Zinzen2010} or protein-protein interactions (proteome) \cite{Lee2002}.
However qualitative information on GRN topology alone is insufficient to understand GRN dynamics.
It has been recognized that quantitative information is required to understand 
the complex dynamical properties of regulatory interactions in living cells \cite{Guet2002,Kitano2002}, 
mainly because dynamics on interaction networks with identical topology still depends 
on the strength of interactions (links) between agents (nodes).
Models of GRN dynamics aid the task of understanding properties of GRN at 
various levels of detail available in experimental data and therefore provide valuable tools for
integrating information from different sources into unifying pictures and for reverse engineering GRN from
experimental data. 
Any model should {\it adequately} reproduce GRN dynamics and {\it sufficiently} exhibit  systemic properties of the GRN, 
including homeostasis,  multi-stability, periodic dynamics, alternating activity, 
self-organized critical dynamics (SOC) and differentiability.

{\it Homeostatic dynamics} regulates the equilibrium concentration levels of agents, e.g. \cite{Semsey2006}, 
{\it multi-stability} shows switching between multiple steady states \cite{Ozbudak2004,Smolen1998}. 
Examples for {\em periodic dynamics} are e.g. the cell-cycle \cite{Lee2002}, circadian-clock \cite{Park2000}, 
I$\kappa$B-N$\kappa$B signaling \cite{Hoffmann2002}, hER dynamics \cite{Metivier2003,Metivier2004} etc. 
Some molecular agents show {\it alternating activity}, i.e. their concentrations alternate between being detectable 
(on) and below detection threshold (off), see e.g. \cite{Metivier2003,Metivier2004}.
{\it Self-organized critical} (SOC) dynamics corresponds to details of 
regulatory dynamics ensuring (approximate) stability within a fluctuating environment through various mechanisms of 
adaptation. 
Finally the property of {\it differentiability} means that cells of multicellular organisms can 
differentiate into various cell-types (liver, muscle, blood, kidney, cancer, ...). The differentiated
cells possess identical GRN but express distinguishable patterns of regulatory activity. The same
GRN therefore can be expressed in different {\it modes} so that some agents become expressed in one mode but not in another \cite{Babu_A}.

Recently it has been reported that both regulation of transcription and mRNA decay rates (i.e. the mRNA turnover)
are necessary to understand experimentally observed expression values \cite{Amorim2010}.
Moreover it has been demonstrated that decay rates of mRNA are cell-type specific \cite{Lee2010}.
Analogously for proteins, where the dominant mechanism is the Ubiquitin driven proteolyse in the Proteasom 
\cite{Ciechanover2005}, protein abundance and therefore their degradation has to be tightly controlled \cite{Babu_C}.  
Also the abundance of proteins and whether certain proteins are produced or not is again cell-type specific \cite{Bossi2009,Bennett2011}. This indicates that decay-rates and their control play a crucial role in cell-differentiation.

Variable decay rates however and the property of differentiability are hardly ever considered in GRN models
where decay rates of agents are usually kept constant. Understanding the effects of changes of decay rates of agents 
therefore is a crucial step towards a deeper understanding of GRN dynamics and the role decay rates play in cell-differentiation.
The GRN is the set of all possible interactions of molecular reactions and bindings. 
The GRN captures all possible features of cells and are responsible for the immense levels of adaptation characteristic to living systems.
What happens when different cell-types express the same GRN in alternative ways?
At any point in time only small subsets of the GRN are active. 
Any active subset of the GRN leads to the expression of particular sets 
of  molecules (expression modes). 
The {\it active regulatory network} at time $t$ is the regulatory sub-network of the GRN, governing the molecular (auto-catalytic) dynamics of all agents which exist 
at time $t$. 
The set of existing molecules forms the {\it active agent set} at time $t$.  
The active network changes over time and typical sequences of active sets represent
what we call the {\it expression modes} of a specific cell-type and their cell-cycle. 
Expression modes themselves can be modified, either locally as a reaction to an external signal, 
or fundamentally through further cell differentiation. 
Active sets of molecules are transient and what is observed in experiments
is a superposition of subsequent active sets, which we call the {\it expressed set of agents} and
the regulatory interactions between the expressed agents the {\it expressed regulatory network}.
To find the property of differentiability in a regulatory network model therefore requires that one network is capable of
producing different expression modes while perturbations (external signal) only modify active sets locally and the 
particular expression mode can be restored.

The six dynamical properties we have listed above have been addressed with a variety of conceptually different models. 
The essence of all these models is that they try to capture the dynamics induced by  
positive and negative feed back loops within the GRN. 
The choice of model depends largely on the type and resolution (coarse graining) of experimental data. 
At the  single cell level  cellular activity  (e.g. concentrations $x_i$ of biochemical agents $i$) can be modeled by 
non-linear (stochastic) differential equations \cite{Turner2004,Bratsun2005} which can 
explain homeostasis, periodic and multi-stable behavior. 
The dynamics governed by a GRN is given by a set of coupled non-linear differential equations
\begin{equation} 
\dot x_i=F_i(x)\,,
\label{NLDE}
\end{equation}
where $F_i$ is a (non-linear) function capturing the GRN. It depends on the vector of concentrations of all the possible $N$ molecular agents in a cell, $x=\{x_i\}_{i=1}^N$.  
$\dot x_i$ is the time derivative of the concentrations $x_i$. Note that $F_i$ can have stochastic components. 
Analysis of such systems is often complicated  by the interplay between fluctuations and non-linearities \cite{Paulsson2000}. 

Differential equation models can be approximated by 
cellular automata, Boolean or piecewise-linear models.
The property of SOC dynamics, or dynamics at the "edge of chaos"  \cite{Langton1990,Mitchell1993,Kauffman1993},
has been studied mainly in the context of cellular automata and Boolean models 
\cite{Bhattacharjya1996,Glass1998,Shmulevich2003}.  SOC dynamics was 
also discussed in continuous differential equation based models \cite{Stokic2008,Hanel2010a}.
Boolean and piecewise-linear models 
share common origins in the work of Glass and Kauffman, \cite{GlassKauffman1973}, and have extensively been used for modeling
and analyzing GRN \cite{Qian2010,deJong2003,Viretta2004,Ropers2005}. 
For their superior properties in approximating non-linear systems (in principle to any suitable precision) piecewise-linear models also are applied in different disciplines, for instance for modeling highly non-linear 
electronic circuits \cite{Rewienski2003}. 

In the context of GRN   
both boolean and piecewise-linear models usually are used for describing non-linear dynamics with switch-like 
regulatory elements frequently observed in biological regulatory processes \cite{Yagil1971,Ptashne1992}. 
Such switches react if the concentration of an agent (the signal) crosses a specific threshold level.
To model such switches in regulation networks of $N$ molecular agents with concentrations $x_i$ 
the space of concentrations $\dom=\{x|x_i\geq 0\}$ is cut into segments defined by the threshold values where
the concentration $x_i$ can trigger a regulatory switch.   
These segments are called {\it regulatory domains} (e.g. \cite{Casey2006}). 
In each such domain Eq. (\ref{NLDE}) gets approximated by a linear equation of the form
\begin{equation}
\dot x_i=\Phi_i+\sum_{j=1}^{N} A_{ij}x_j\,,
\label{linearisation}
\end{equation}
where the $\Phi_i>0$ 
are production rates and $A_{ij}$ are interaction matrices between agents.
If $A_{ij}>0$, then $j$ promotes the production of $i$. If $A_{ij}<0$, then $j$ suppresses $i$. If $A_{ij}=0$ $j$ has no influence
on $i$. The diagonal elements $A_{ii}<0$ are {\it decay rates}, $D_i=-A_{ii}$. 
Non-linear effects purely come from concentration $x_i$ passing threshold levels, where the dynamics of $x$ switches from one to another
regulatory domain with different values of $\Phi$ and $A_{ij}$. 
Equation (\ref{linearisation}) is a slight generalization of the 
Glass-Kauffman PLM, \cite{GlassKauffman1973,Casey2006}, where $A_{ij}=0$ except for the (usually) fixed 
decay rates $D_i$, so that only the production rates change with the regulatory domain. 

Given that the interaction matrix $A$ of the regulatory network is invertible 
(which is almost certainly true for the biologically relevant range of connectivities of GRN)  
Eq. (\ref{linearisation}) can be rewritten 
\begin{equation}
\dot x_i=\sum_{j=1}^N A_{ij}\left(x_j-x_j^*\right)\,,
\label{eq:rate-eqn}
\end{equation}
with 
$x^*$ 
being the solution of the equation 
$\Phi_i=-\sum_jA_{ij}x_j^*$.  
The fixed-point $x^*$ is stable (unstable) and 
$x_i$ will be attracted (repelled) by $x_i^*$. If $x^*$ is stable and $x^*_i>0$ for all 
$i$ then $x(t)=x^*$ is a stationary solution of Eq. (\ref{linearisation}).

Not all models approximating nonlinear differential equation descriptions of GRN are equally suited 
to capture all GRN properties discussed above simultaneously depending on whether 
discrete (Boolean, cellular automata) or smooth (differential equation) features dominate the model. 
However there exists a surprisingly simple class of models which exhibits 
{\it all} desired GRN properties.

Here we present such a simple model that captures all of the above dynamical properties.
We find that the alternating dynamics plays a key role for the stability of regulatory systems and for the formation of SOC dynamics
in particular \cite{Stokic2008,Hanel2010a}.
Most importantly we are able to show that even unspecific control over decay rates, changing the magnitude of all decay rates simultaneously by a (small) factor, leads to 
"cell differentiation", i.e. the same regulatory network enters different expression modes, displaying different sequences
of active regulatory networks.  

We show that experimental facts, linking variations of decay rates observed between different cell-types of an organism
to variations of the abundance of intra-cellular biochemical agents in these cell-types,
correspond to (a) differences in the {\it expressed} genetic regulatory network, and
(b) these differences can be controlled via decay rates of intracellular agents. 
In other words typical expression modes (cyclical sequences of successive active sub-networks of the GRN) can be altered and switched 
by controlling decay rates.

\section*{The model}

Glass-Kauffman systems, \cite{GlassKauffman1973}, 
produce positive concentrations $x_i(t)>0$ for all times $t$ given positive initial conditions $x_i(0)>0$. 
This however makes it impossible to produce alternating activity of agents since zero-concentrations $x_i(t)=0$ can not appear.
Therefore we have to generalize Glass-Kauffman systems to more general forms of invertible interaction matrices 
$A_{ij}$ where the positivity of solutions of Eq. (\ref{linearisation}) is not implicitly guaranteed, but where 
positivity (non-negativity) is ensured as a constraint to the system, 
\begin{equation}
	x_i(t)\geq0 \quad \forall\ \mbox{agents}\ i, \mbox{and times}\ t \quad .
	\label{eq:PC}
\end{equation}
This constraint alters the linear dynamics of Eq. (\ref{linearisation}) in the following way.
Whenever a concentration $x_i$ becomes zero at time $t$ then $x_i(t')$ remains zero
for $t'>t$ for as long as $\dot x_i(t')<0$, according to Eq. (\ref{linearisation}). 
If $\dot x_i(t''')\geq 0$ for $t'''\geq t''>t$ then $x_i(t''')$ is no longer subject
to the positivity constraint and continues to evolve according to Eq. (\ref{linearisation}) again.
Agent $i$ is said to be {\it active} at time $t$, if $x_i(t)>0$ and {\it inactive}, if $x_i(t)=0$.

The positivity constraint Eq. (\ref{eq:PC}) implies the following consequences. 
At any point in time there will be a sub-set of agents with non-vanishing concentrations which we call 
the {\em active set} of agents. The remaining agents  have zero concentration, 
and therefore do not actively influence the concentrations of any of the non-vanishing agents. 
There exist $2^N$ different active sets, i.e.  $2^N$ combinations in which $N$ agents can be active or inactive.
Each active set can be uniquely identified by an index $s=1,\dots,2^N$.
In the course of time $t$ some agents will vanish while others  re-appear, so that one effectively observes a sequence of sets of active agents  
\begin{equation}
s_0\stackrel{t^{\rm switch}_1}{\longrightarrow} s_1\stackrel{t^{\rm switch}_2}{\longrightarrow} s_2\stackrel{t^{\rm switch}_3}{\longrightarrow} \dots\,, 
\end{equation}
$s_0$ being the initial active set.  The active set $s_{m-1}$ switches to active set $s_m$ at time $t^{\rm switch}_m$.
In each time interval $T_m=\left[t^{\rm switch}_m\,t^{\rm switch}_{m+1}\right]$ 
of duration $\tau_m=t^{\rm switch}_{m+1}-t^{\rm switch}_{m}$ 
it is thus  possible to only consider the regulatory sub-network acting on the set of active agents $s_m$. 
This sub-network is described by the part of the full interaction matrix $A_{ij}$, where $i$ and $j$ 
are restricted to the set of active agents $s_m$. 
These sub-matrices we call {\em active networks} and denote them by $A_{\rm act}^{s_m}$. 
The concentration vector of active agents we call $x_{\rm act}^{s_m}$.
Active agents also "feel" a modified {\it effective} fixed point $x^{*\,s_m}_{\rm act}$,  such that finally for $t\in T_m$ the 
concentrations of the active agents follow a linear equation 
\begin{equation}
\dot x_{{\rm act},\,i}^{s_m}(t)=\sum_{j\,{\rm is}\, {\rm active}} A_{{\rm act},\,ij}^{s_m}
\left(x_{{\rm act},\,j}^{s_m}(t)-x_{{\rm act},\,j}^{*\,s_m}\right)\,.
\label{eq:sub-rate-eqn}
\end{equation}
We refer to such systems as {\it sequentially linear} systems. 
The attractiveness of this description arises through the fact that it becomes possible to 
understand the dynamics by considering the sequences of active networks
\begin{equation}
A_{\rm act}^{s_0}\stackrel{t^{\rm switch}_1}{\longrightarrow} A_{\rm act}^{s_1}\stackrel{t^{\rm switch}_2}{\longrightarrow} A_{\rm act}^{s_2}\stackrel{t^{\rm switch}_3}{\longrightarrow} \dots\,, 
\end{equation}
which allows to analyze dynamical properties in terms of eigenvalues and eigenvectors of the active sub-matrices $A_{\rm act}^{s_m}$ (see Materials and Methods). 
This model can be shown to be mathematically equivalent to  \cite{Stokic2008,Hanel2010a}.

\subsection*{Cell differentiation in the sequentially linear dynamics}
$\,$
In the picture of sequentially linear dynamics it becomes possible to identify operational modes of a cell as a particular sequence of active networks.    
Cell types in ordinary operational modes may be classified by specific sequences. 
As a hypothetical  example  a liver cell under typical conditions might be characterized by a {\it periodic} sequence $A_{\rm act}^{9}\to A_{\rm act}^{10}\to A_{\rm act}^{46}\to A_{\rm act}^{2}\to A_{\rm act}^{9}$, 
whereas an endothelial cell is given by $A_{\rm act}^{123}\to A_{\rm act}^{2}\to A_{\rm act}^{4}\to A_{\rm act}^{209}\to A_{\rm act}^{9}\to A_{\rm act}^{77}\to A_{\rm act}^{123}$. 
Note that all types share the same full regulatory network $A$.
This separates timescales of the dynamics: on the fast timescale the dynamics is
continuous and characterized by linear changes of the concentrations $x_i$. 
On the slower time-scale the dynamics is characterized by discrete changes of active sets.
The change from one sequence of active sets to another can be interpreted as the expression 
modes of different cell-types (cell differentiation) and we show that changes in decay rates of molecular species trigger 
switches between expression modes.  

\subsection*{Example}

As an example for sequentially linear dynamics we consider a system with $N=4$ molecular agents,  
$x_i^*=100$, $D_i=-A_{ii}=0.23$ for all agents $i=1,\dots,4$, and  a regulatory network  given by 
{\small
\begin{equation}
A= \left( \begin{array}{lllll}
 -0.23 & -0.1  & \ \ 0    & \ \ 0.1   \\
 \ \ 0   & -0.23 & \ \ 0.2  & \ \ 0   \\
 \ \ 1   &\ \ 0	   &  -0.23 &  -1  \\
-0.8   & -0.8  & \ \ 0.1  &  -0.23  \\
  \end{array} \right)\,.
\label{eq:matrix}
\end{equation}
}
%
%
 \begin{figure}[!ht]
 \begin{center}
 	\includegraphics[width=4in]{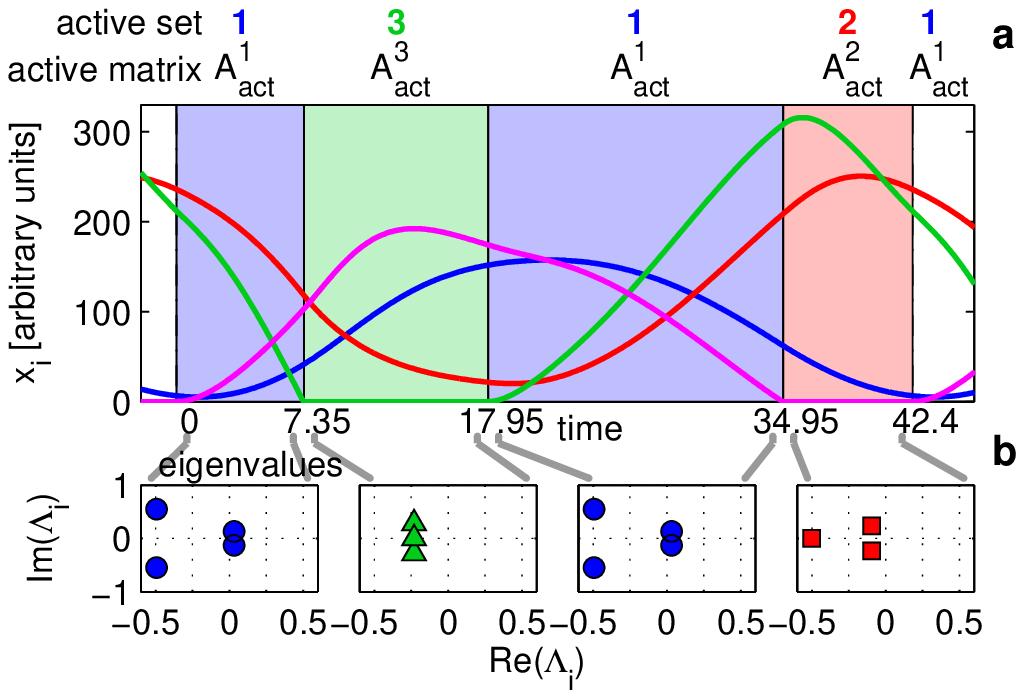} \\  	 
 	\includegraphics[width=2.4in]{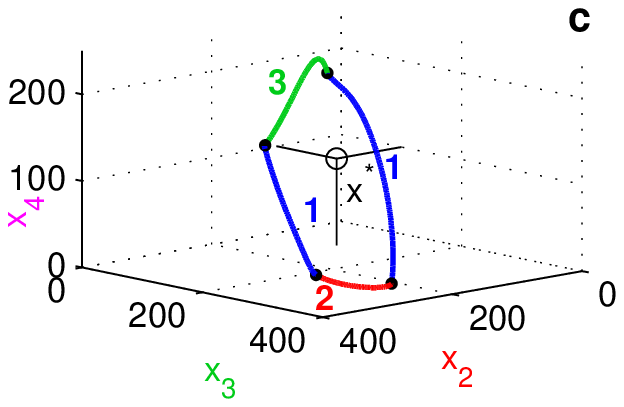}	 
 	\includegraphics[width=1.6in]{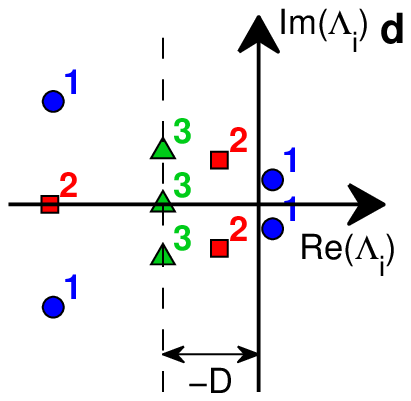}%
 \end{center}
 	\caption{ {\bf Periodic dynamics and active sets}. Sequentially linear system with decay rate $D_i=0.23$ and the fixed point $x^*_i=100$ for all agents $i$
 		simulated with time-increment $dt=0.05$. Periodic time-series organized into a sequence of four domains with three 
 		different active sets. 
 		For each time-domain 
 		the associated spectrum of eigenvalues for the active sets 
 		is shown in (b). 
 		In (c) a 3d Poincare map of the 
 		limit cycle is plotted together with the projection of $x^*$ in the center. 
 		The domains are marked with bold numbers and 
 		switching events with dots. 
 		(d) The eigenvalue spectra of the different subsystems are plotted in the imaginary plane.  
 		The shift of the spectrum along the real axis depending on the decay rate $D$ is indicated. 
		\label{figure_main1}	}
\end{figure}
%
The dynamics of this system (over one period) is shown in Fig. \ref{figure_main1} a. 
The property describing the stability of an active set $s_m$ is the maximal real part of 
the eigenvalues $\Lambda_{\rm act}^{s_m}$ of the active matrix $A_{\rm act}^{s_m}$ 
denoted $L_{\rm act}^{s_m}=\max\mbox{Re}(\Lambda_{\rm act}^{s_m})$. The number $q$ denotes
the number of time-domains in a periodic sequence of active networks and $z$ is the number of
different sub-networks that are activated in a sequence 
(see also \SI).
In this example there are four time-domains ($q=4$) associated with three different active sets ($z=3$) %
which are periodically repeated. 
The sequence starts in time-domain $1$ with active set $s=1$ with maximum real eigenvalue $L_{\rm act}^1= 0.03$. 
Positive $L_{\rm act}^1$ means that the fixed point of the active set is unstable and the associated leading eigenvalue implies
that the concentration of one agent (green) is decaying to zero. 
The positivity condition deactivates this agent as its concentration becomes zero 
and the system enters time-domain $2$ as the active set
switches to $s=3$ with $L_{\rm act}^2=-0.24$. Negative $L_{\rm act}^2$  means that the fixed point $x_{\rm act}^{*\,3}$ 
is stable and $x_{\rm act}^2$ tries to approach $x_{\rm act}^{*\,2}$. This leads to the deactivated 
agent (green) becoming produced again and the system switches back to $s=1$ entering the third time-domain.
In time-domain $3$ the initial conditions differs from the one in time-domain $1$ and
a different node (magenta) gets deactivated. The system switches to $s=2$ 
with $L_{\rm act}^3=-0.09$ at the beginning of the fourth time-domain. 
This means $x_{\rm act}^{*\,2}$ is a stable fixed-point and 
the inactive node (magenta) eventually gets produced again as the system switches back to the beginning ($s=1$) 
and enters the next period.
The system is thus precisely characterized by the 
sequence $A_{\rm act}^1\to A_{\rm act}^3\to A_{\rm act}^1\to A_{\rm act}^2\to A_{\rm act}^1$.
The eigenvalue spectra of the sub-matrices $A_{\rm act}^{s_m}$ associated with subsequent time-domains $T_m$ are shown in 
Fig. \ref{figure_main1} b.
Fig. \ref{figure_main1} c shows a projection of the trajectory into a three dimensional Poincare map.
Fig. \ref{figure_main1} d shows the eigenvalue spectra of the different active sub-systems of the dynamics.

\begin{table}[ht]
\begin{center}
	\begin{tabular}{@{\vrule height 10pt depth4pt  width0pt} ccccrr}	
	 	time-domain & $s$ & $N_{\rm on}$ & $\tau_s$ &  
$L_{\rm act}^s$ & stability \\ 
	 	1 & 1 & 4 & 7.35 &  0.033  & \mbox{unstable} \\
	 	2 & 3 & 3 & 10.6 &  -0.24 & \mbox{stable}  \\
	 	3 & 1 & 4 & 17.0 &  0.033  & \mbox{unstable}\\ 
	 	4 & 2 & 3 & 7.45 & -0.094  & \mbox{stable}  \\
	\end{tabular}
\end{center}
\caption{{\bf Properties of 4-node example system}.
 Some characteristics of the four node system shown in Fig \ref{figure_main1} are listed, 
	including the index of the time domain, the index of the sub-system $s$, the number of active nodes 	$N_{\rm on}$, the time the system spends in the $s$'th sub-system, the real-part of the leading 
	eigenvalue of $s$, and whether sub-system $s$ is stable or not.  
} 
\label{table_1}
\end{table}
%
%

Some details of the dynamics, like the existence of multiple stable fixed-points, the 
periodicity of bounded attractors and temporal self-organization, can be mathematically fully understood.
In \cite{Stokic2008,Hanel2010a} it was already shown mathematically that sequentially linear models exhibit {\it homeostasis}, 
{\it multi-stability}. This has been demonstrated for a wide range of system size $N$, and a number of interactions (connectivity)
and fixed decay rates. 
{\it Periodic dynamics}, and {\it self-organized critical} dynamics have been noted
in  \cite{Stokic2008,Hanel2010a} but were not clarified and require further explanation which is given in detail 
in the 
\SI, where also a simple temporal balance condition is described and derived.

The temporal balance condition states that the time-average over the real parts of the leading eigenvalues $L_{\rm act}^{s_m}$ 
of the matrices $A_{\rm act}^{s_m}$ in a sequence of active networks approximate the Lyapunov exponent $\lambda$.
The Lyapunov exponent $\lambda$ measures the overall stability of a system ($\lambda<0$ stable, $\lambda>0$ instable, $\lambda=0$ critical) and for sequences following a periodic attractor $\lambda$ can be shown to be exactly zero.
Inserting the values for $\tau_s$ and $L_{\rm act}^s$ from table 
(\ref{table_1}) into the balance condition, Eq. (\balanceeq) gives the value $-0.055$ as an approximation of $\lambda$ 
(which has an exact value of zero). Although the balance equation gives only a crude approximation of the Lyapunov exponent 
it allows to understand why the example-system spends more time in the weakly instable time-domain $1$ and $3$, 
than in the stable time-domains $2$ and $4$ which is obviously true from Fig. \ref{figure_main1}. Strong convergence needs less
time to compensate for weak divergence.   

Temporal balance is a consequence of the mechanism of self-organization that fine-tunes switching times 
such that stable parts of the dynamics compensate instable parts of the dynamics exactly. This mechanism can be understood in the following way.
Sequentially linear systems try to converge to a fixed point. If it is reached the system becomes static. 
The fixed point might not be {\it accessible} however, meaning that the trajectory on the way toward the fixed point hits 
a boundary (Fig. \ref{figure_main1} c) causing a switching event which changes the dynamics so that the system 
now is attracted by a different effective fixed point, 
which it tries to reach. 
If the system does not converge to an accessible fixed point it 
is either unstable and some concentrations $x_i$ diverge, or the system circles through some of the 
$2^N$ possible active sets and converges onto an effective attractor - characterized in the sequence of active networks.  
In the later case small perturbations of $x(t)$ on the attractor will vanish with time.
This allows to show that bounded dynamics that does not converge to a fixed-point has to be periodic 
(\SI).
Switching times are not static but react to perturbations 
of concentrations $x_i$. Perturbations shift the occurrence of switching times proportional to the magnitude of the perturbation. 
This has the effect that switching events act like sliding "focal planes" allowing 
the perturbed dynamics to "refocus" onto the periodic attractor.     
While the perturbed dynamics returns to the attractor switching times cumulate small time-shifts 
resulting in a phase-shift of the periodic dynamics. 
A perturbation is remembered as a phase-shift of the periodic dynamics which neither grows exponentially nor dies out.
The Lyapunov exponent therefore is zero and the systems self-organizes to the "edge of chaos" by adaptation of switching times.  
Stable adaptive dynamics is a result of this "temporal self-organization".

 \begin{figure}[!ht]
\begin{center}
	 \includegraphics[width=4in]{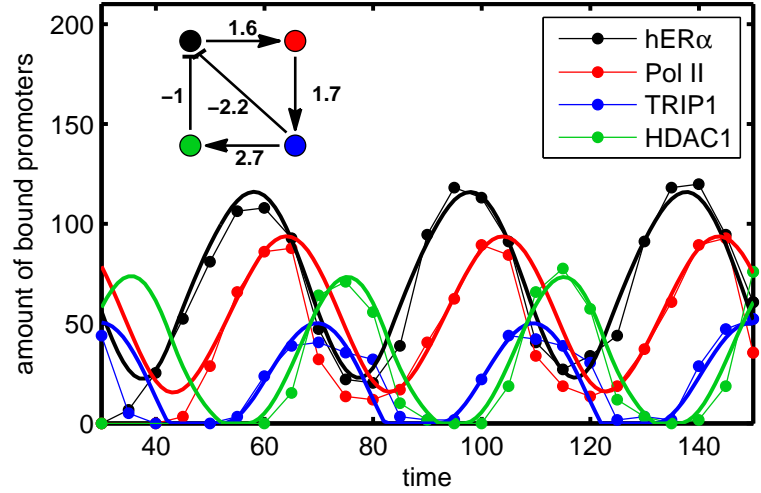}%
\end{center}
	 \caption{ {\bf Adequacy of sequentially linear systems}. Time series of periodic binding of four proteins to the pS2 promoter
	 after addition of estradiol - experimental data has been extracted from \cite{Pigolotti2009}, where a negative feedback-loop was
	 proposed to explain the dynamics. Experimental data due to \cite{Metivier2003} and \cite{Metivier2004} (dotted lines) is compared
	 with a simulation of a SL system, based on the network shown in the inset, with uniform decay rates $D_i=1.08$ for all agents and fixed point 
	 concentrations $ x^*=\left[ 75 ; 60 ; 20 ; 30\right] $. Correlation coefficients for simulated and measured time-series are 
	 $ C_i=( 0.97; 0.84; 0.94 ;0.97 ) $ for time larger $40$ and agents $i$ in order of the legend. 
	 The model simulation uses zero concentrations for all agents as initial condition and a time increment $dt=0.1$. 
	 For matching the simulation with experiment time in the model is shifted by $-40$. 
	  \label{figure_main2} }
 \end{figure}

\section*{Results\label{sec_results}}
We first show that the model is able to explain actual empirical data, including alternating dynamics. 
Figure \ref{figure_main2} shows data of molecular concentrations $x_i(t)$  (hER$\alpha$ (black), Pol II (red), TRIP1 (blue), HDAC1 (green)) over three periods of about 40 minutes time. These four agents are all part of the 
human estrogen nuclear receptor dynamics. The source of the Data is Metivier et. al. \cite{Metivier2003}. Data points were taken from  Pigolotti et al. \cite{Pigolotti2009} and 
the actual values of the matrix elements  
{\small
\begin{equation}
A= \left( \begin{array}{lllll}
   -1.08 & \ \ 1.6 	& \ \ 0  	 & \ \ 0   \\
\ \ 0  	 &    -1.08 & \ \ 1.7  & \ \ 0   \\
   -2.2  & \ \ 0	 	&	   -1.08 & \ \ 2.7   \\
   -1	 	 & \ \ 0	 	& \ \ 0.1	 &  	-1.08  \\
  \end{array} \right)
\label{eq:matrix_hER}
\end{equation}
}
are bests fits with identical decay rates for optimal explanation of the data.
The TRIP1 data (blue) shows {\it alternating activity} which is reproduced perfectly by our sequential linear model.

\subsection*{Decay rates and expression modes}      
In the following we show how the change of decay rates induces changes from one 
cell-type to another. In particular we show how changes of the overall 
strength of the decay rates results in differentiated dynamics, i.e. in distinct sequences of active expressed networks.
This allows to understand recent experimental observations which indicate correlations between
cell-type, expressed sets of agents, and decay-rates \cite{Babu_A,Amorim2010,Lee2010,Babu_C,Bossi2009,Bennett2011}. 

For a fixed interaction network temporal self-organization can be maintained for a wide range of decay rates $D$.  
We show this in the same $4$-node system considered in Fig. \ref{figure_main1} by only varying the decay rate 
$D=-A_{ii}$ from Eq. (\ref{eq:matrix}). 
\begin{figure}[!ht]
\begin{center}
	\includegraphics[width=4in]{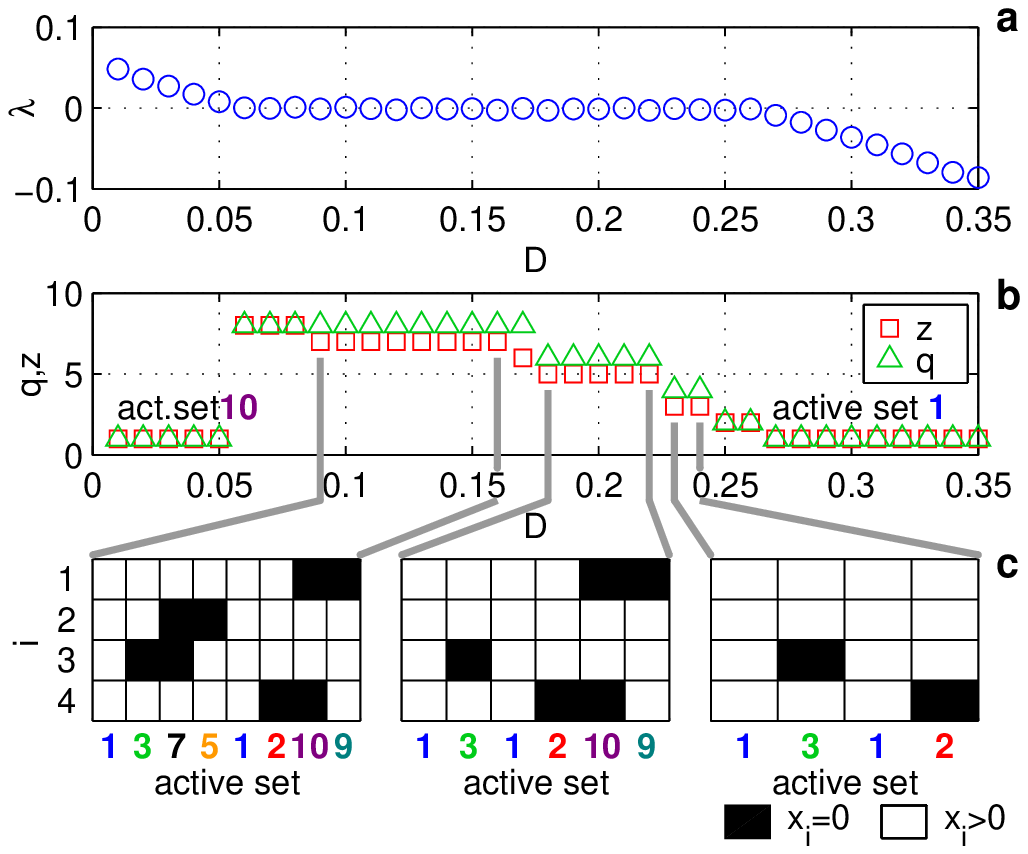} \\ %
	\includegraphics[width=2.4in]{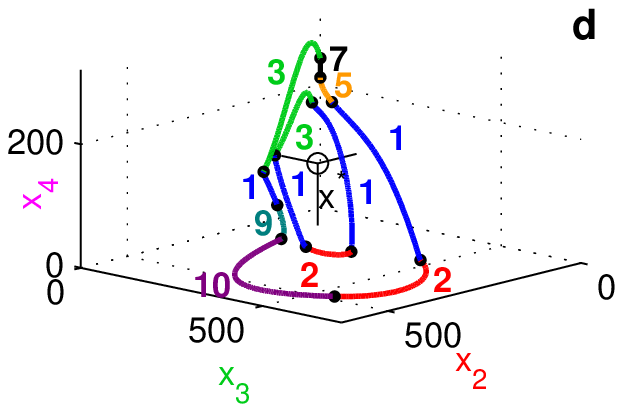}
	\includegraphics[width=1.6in]{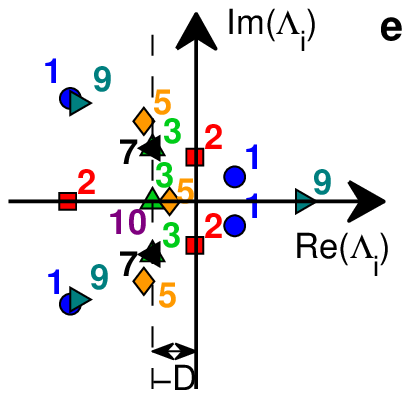}%
\end{center}
	\caption{ {\bf The edge of chaos}. The Lyapunov exponent $\lambda$ of the four node system, 
		Eq. (\ref{eq:matrix}), is shown in (a) as a function of 
		the decay rate $D$, which exhibits a "plateau" with $\lambda=0$ in the range $0.06<D<0.26$.	
		In (b) the length $q$ of the periodic sequence of domains is 
		plotted in green triangles and the number of different active sets $z$ as red squares. 
		In (c) the sequences of active sets are shown for decay rates $D=0.23$, $=0.2$ and $0.14$. 
		The limit circles for decay rates $D=0.23$ (short sequence) and $D=0.14$ (long sequence) are visualized in 
		(d) in a Poincare map using three out of four phase-space dimensions. 
		With decreasing $D$ the radius of the limit circle becomes wider and 
		additional sets (marked with colors) become active. 
		In (e) the spectra of eigenvalues are shown for all the appearing active sets with $D=0.14$. 
	\label{figure_main3}	}
\end{figure}
Figure \ref{figure_main3} a shows the Lyapunov exponent $\lambda$ as a function of $D$. A plateau, where
$\lambda\sim0$, is clearly visible. If the decay rate is larger than a critical value $D>0.26$, the Lyapunov exponent 
becomes negative ($\lambda<0$) and the system stable.
If the decay rate is smaller than a critical value of  $D<0.06$, temporal balance can not be achieved any more, refocusing breaks down, and the system becomes chaotic and trajectories diverge exponentially with $\lambda>0$. In Fig. \ref{figure_main3} b the 
length of the periodic sequences $q$ (green triangles), which is the number of time-domains in a sequence, 
and the number $z$ of different active sets activated in this sequence
(red squares) is depicted. 
Figure \ref{figure_main3} b also shows that at several critical values of 
$D\sim 0.088,\,0.162,\,0.171,\,0.224,\,0.246,\,0.263$ in the plateau region the sequences of active regulatory sub-networks changes when temporal balance can no longer be established merely 
by adapting the switching times of a sequence.
Sequences do not usually change completely at critical values of $D$ and are only expanded by additional active subsets. 
This can be seen clearly in the 3D Poincare map of the dynamics Fig. \ref{figure_main3} d, 
where the sequence of subsystems $s$ given by $1\to 2\to 1\to 3\to 1$  (for $D=0.23$) gets 
expanded to the sequence 
${\bf 1}\to {\bf 2}\to 10\to 9\to {\bf 1}\to  {\bf 3}\to 7\to 5 \to {\bf 1}$ (for $D=0.14$).
In the \SI, Fig. (1), the longer sequence is also shown in the space of all possible active sets.   
The mathematical reason why such critical decay rates exist is that changes of $D$ shift the eigenvalue spectra of the active 
interaction matrix $A_{\rm act}^s$, shown in Fig. \ref{figure_main3} e, 
along the real axis. The real part of the leading eigenvalues, $L_{\rm act}^s$,
is becoming smaller (larger) than zero and $x^{*\,s}_{\rm act}$ becomes 
an attractor (repellor) of $x_{\rm act}^s$. The stable fixed point then either is accessible and the dynamic 
changes from periodic to stationary or inaccessible and the dynamic changes qualitatively but remains periodic.
Which agents become active in a given active set $s$ is depicted in Fig. \ref{figure_main3} b 
for three different sequences of active sets associated with three different ranges of the decay 
rate $D$ indicated by gray lines. If node $i$ is active in active set $s$
then the associated field is white and black otherwise.

 \begin{figure}[t]
\begin{center}
	 \includegraphics[width=4in]{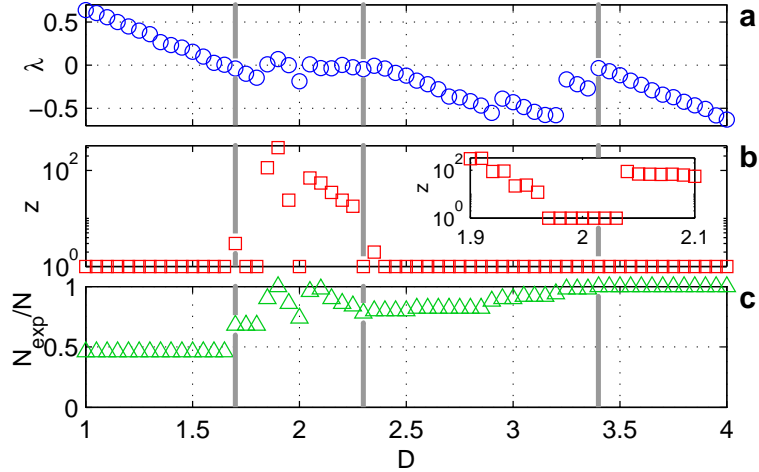}%
\end{center}
	 \caption{ {\bf Degradation rates and active networks}. Example of a SL system with $N=50$ and $\mk=10$ and identical initial conditions for all values of $D$.  
	 (a) The Lyapunov exponent, (b) the number of 
	 active sets $z$ in a period (if $z=1$ then the the sequence is not periodic but a 
	 steady state!), and
	 (c) the fraction of expressed nodes are plotted as functions of the uniform decay rates $D_i=D$.
	 For $D>3.4$ $x^*$ is stable. In the range $2.4<D<3.4$ the $x^*$ has become unstable but 
	 the plateau ($\lambda=0$) can not form since the dynamic finds active sets $s$ with stable 
	 and accessible $x^{*\,s}_{\rm act}$. 
	 The inset in (b) shows that in the plateau region a small window, $1.97<D<2.03$, 
	 exists where again an active set $s$ contains an accessible $x^{*\,s}_{\rm act}$ attracting  
	 the dynamics.   
	 In the range $1.7<D<2.4$ the plateau forms and dynamics gets periodic. For $D<1.7$ the system 
	 gets unstable.
	  \label{figure_main4} }
 \end{figure}
%

The number of {\it expressed} agents $N_{\rm exp}$ is the number of agents that are 
active at least once during a period of the dynamics. 
To demonstrate that not only the periodic activation of agents depends on $D$
but also the number of expressed nodes $N_{\rm exp}$ itself,  
we consider a larger sequentially linear system with $N=50$ agents. 
The interaction matrix of the system is a random matrix with average connectivity $\langle k\rangle=10$, 
meaning for each node $10$ interactions with other agents have been randomly chosen with equal probability.
Each non-zero entry, describing such an interaction, is drawn from a normal distribution with mean zero and a standard deviation of $\sigma=1$. This means that the interaction strength is of magnitude $1$ on average and has positive or negative sign with equal probability.
In Fig. \ref{figure_main4} a the Lyapunov exponent $\lambda$, 
in Fig. \ref{figure_main4} b the number $z$ of sets that become active during a cycle 
and in Fig. \ref{figure_main4} c the fraction of expressed agents $N_{\rm exp}/N$ is plotted as 
a function of $D$. 
For large decay rates ($D>3.4$) the system is stable and $x^*$ is a fixed-point of the dynamics.  
As $D$ decreases $x^*$ becomes unstable for $D\sim 3.4$. However for $2.3<D<3.4$ the system  ends up in some 
stable accessible fixed point $x^{*\,s}_{\rm act}$ so that $x(t)$ approaches  a stationary state and $z=1$. 
In this range $N_{\rm exp}$ increases  with $D$. The $\lambda\sim 0$ plateau with stable self-organized critical 
dynamics ($z>1$) only emerges in the range $1.7<D<2.4$ where number of active sets $z$ and expressed network size $N_{\rm exp}/N$ vary strongly. $N_{\rm exp}/N$ varies between $1$ and $0.5$ which means that changes of the decay rate can 
induce changes of the size of the expressed network comparable to the magnitude of the full interaction network. 
A small window of stability exists for $1.97<D<2.03$ (see inset).    

The strong dependence of $N_{\rm exp}/N$ on the decay-rate $D$ (up to $50\%$ of the total regulatory network)
demonstrates clearly that decay-rates alone massively influence sequences of active systems without changing 
the interaction strength between agents in the regulatory network at all.
Moreover, decay rates can also cause switches between fixed-point dynamics and periodic dynamics. 
While fixed points favor larger decay-rates (in the example $D>2.3$) there can also exist fixed points for smaller decay rates
(window of stability $1.97<D<2.03$) where systems favor periodic dynamics.

\section*{Discussion\label{sec_discussion}}

We presented a model which de-composes the dynamics of molecular concentrations --  governed by the full molecular regulatory networks --  into a temporal sequence of active sub-networks. This novel type of model allows not only to reduce the vast complexity of the full  regulatory network into sub-networks of managable size but further  to approximate the complicated dynamics by linear methods. The intrinsic non-linearities in the system which lead to  alternating  dynamics in concentrations (as found in countless experiments) are absorbed into switching events, where the dynamics of one linear system switches to another one. 
In this view different cell types correspond to different sequences of active sub-networks over time. 

These sequentially linear models allow not only for the first time to describe all the relevant dynamical features of the GNR  (homeostasis, multi-stability, periodic dynamics, alternating activity, differentiability, and self-organized criticality), but also offers the handle to understand the role of molecular decay rates. 
The fact that sequentially linear dynamics properly models homeostasis, multi-stability and periodic behavior was shown in  \cite{Stokic2008,Hanel2010a}. Here we have shown how self-organized criticality (Lyapunov exponent self-regulates to zero) 
arises as a consequence of temporal balance of switching events. 
This  requires agents to show alternating activity (being repeatedly on and off), which is a natural  property by construction of sequentially linear models, and which has posed an unresolved problem of  previous models such as the Glass-Kauffman \cite{GlassKauffman1973} model and its many variants. 
The mechanism behind self-organized criticality  is based on adaptive switching times which effectively  lead to refocusing of perturbed dynamics onto the attractor of sequences of active sub-networks. Such a  temporal self-organization causes long time memory of perturbations in terms of phase-shifts of the otherwise unchanged periodic dynamics, causing the Lyapunov exponent to become zero. 
In other words slight perturbations, e.g. noise, only cause time-shifts of the sequence of regulatory reactions  but do not change the underlying sequence. 
Perturbations  are "remembered" by the system by non vanishing phase-shifts and the dynamics gets "refocused" onto the periodic attractor merely accumulating a time-shift. This has the consequence that the Lyapunov exponent is zero and the system self-organizes its criticality by adapting switching-times.
Practically this means that a system  balances the time it spends in its active sub networks with  stable and unstable dynamics (temporal balance).

Applying the sequentially linear model  to the problem  of cell-differentiation we demonstrate that different levels of decay rates are one to one related with transitions from one active sub-network sequence (cell type) to another. 
This might be a key ingredient to understand a series of recent experimental facts reported on the role of decay-rate regulation systems and the role of noise in cell differentiation \cite{Babu_A,Amorim2010,Lee2010,Babu_C,Bossi2009,Bennett2011}. 
We  found that by varying  the decay rates only, while keeping the complete regulatory network fixed over time, substantially modifies the temporal organization of regulatory events. In particular the decay rate controls the number of expressed agents, the sequence of active sub-networks, and sometimes even the type of solution (stable, stationary, periodic). 
The changes occur at critical levels of decay rates and changes can be drastic. For example we find situations where a 5\% variation of the decay rate causes an approximate doubling of the number of expressed agents.  
This demonstrates that different expression modes, which distinguish different cell-types from each other, can be very efficiently obtained by controlling the decay rates of agents without altering any interactions between agents in the regulatory network, which is very costly in an evolutionary sense. 
These findings highlight the importance of intracellular decay rate control mechanisms and the role of noise in cell differentiation. 

%
%
%

\section*{Materials and Methods}

\begin{figure}[t]
\begin{center}
	\includegraphics[width=4in]{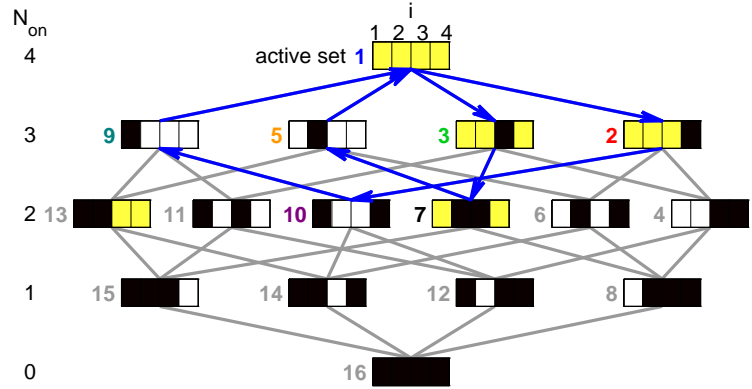}%
\end{center}
	\caption{{\bf Tree of active sets}. Tree of all existing active sets $s$ for system shown in article Fig. (3). 
		In set $1$ all $x_i>0$, yellow background stand for complex leading eigenvalues of the active 
		interaction matrix. 
		Black indicates that the agent associated with that index is not active. 
		The gray lines indicate to all possible switching events where the number of active agents 
		$N_{\rm act}$  changes $\pm1$. Blue arrows mark the observed sequence of the dynamics for the 
		examples Eq. (\eqmatrix) with $D=-A_{ii}=0.14$.
\label{fig-systree} }
\end{figure}

\subsection*{Eigenvalues}
The eigenvalues $\Lambda\in\mathbb{C}$ and eigenvectors $v$ of a matrix $A$ are defined as solutions of the matrix equation
$\Lambda v=Av$. The solution of a linear differential equation $\dot x= A(x-x^*)$ is of the form $x(t)-x^*=\exp(At)(x(0)-x^*)$.
For large times the $x(t)$ will therefore point into the direction of the eigenvector $v_1$ with the
eigenvalue $\Lambda_1$ with the largest real part and $(x(t)-x^*)\sim \exp(\Lambda_1 t)v_1$ as $t$ gets large. 
If the largest real part of $\Lambda_1$ is larger (smaller) than zero $|x(t)-x^*|$ will grow (decay) exponentially and $x^*$ is an unstable (stable) fixed point of the differential equation.

\subsection*{Fixed points and attractors}
Let $L_{\rm act}^{s}$ be the maximal real part of the leading eigenvalue 
of the active interaction matrix $A_{\rm act}^s$ associated with the active subset $s$. 
The effective fixed point $x^{*\,s}_{\rm act}$  is {\it stable} and perturbations of concentrations
vanish if $L_{\rm act}^s<0$. The fixed point is {\it accessible} if $x_{\rm act}^s$ approaching $x^{*\,s}_{\rm act}$
does not cause a switching event and {\it inaccessible} otherwise. 
Stationary solutions of a sequentially linear system therefore require fixed points that are both stable and accessible.

\subsection*{Periodicity of attractors and self-organized criticality}

Suppose a bounded attractor exists for a sequentially linear system with $N$ agents $i$. 
The perturbation $x(t)\to x'(t)$ at time $t=t_0$ also effects later switching times of agents $i$, i.e. $t_m\to t'_m$ such that 
$|\tau'_m-\tau_m|<C |\delta x_m|$ for some constant $C>0$, where $\delta x_m=x'(t'_m)-x(t_m)$. 
Since $|\delta x_m|\to 0$ sufficiently fast as $m\to\infty$ (there exists an attractor)  
the cumulated time shift $t'_m-t_m$ of switching times remains finite for all times. This shows that the 
perturbed $x'$ behaves (after some time) just like the unperturbed $x$ only  shifted in time. 
Perturbation neither vanishes nor grow exponentially, 
and the Lyapunov exponent can only be zero ($\lambda=0$).
Moreover, since the number of active sets is finite ($2^N$) and the dynamics is bounded the
concentrations have to return to values on the attractor with arbitrary precision 
within some finite return-time. The remaining concentration difference can be seen as a perturbation
so that the attractor can only be a periodic cycle. The time-shift produces a phase-shift
of the periodic dynamics.

\subsection*{Stability: the maximum Lyaponov exponent}
While eigenvalues tell us something about the stability of a fixed point the
Lyaponov exponent $\lambda$ tells something about the stability of the dynamics $x(t)$ itself.
The Lyapunov exponent \newline
$\lambda=\lim_{t\to\infty}\log(|\delta x(t)|)/\log(|\delta x(0)|)$ 
measures how a small perturbation $\delta x(t)$ grows with time. If $\lambda<0$ the perturbation vanishes exponentially with time 
or grow exponentially if $\lambda>0$. System with $\lambda>0$ are chaotic (in-stable dynamics extremely
sensitive to noise or perturbations) while $\lambda<0$ indicates stable dynamics insensitive to perturbations and noise.
Systems with $\lambda=0$ are special as their dynamics is sensitive to noise and perturbations without "overreacting"
like chaotic systems. These systems at the "edge of chaos" adapt to fluctuations but 
remain close to their unperturbed dynamics.  

\subsection*{Temporal self-organization of switching events} 
Here we derive a simple approximation of the Lyapunov exponent of sequentially linear dynamics which explains
temporal self-organization quantitatively. This is necessary for understanding why switching in general happens 
between active networks with stable and unstable dynamics and not from one stable stable (unstable) to another
stable (unstable) active network.  

Qualitative analysis of bounded attractors of sequentially linear dynamics 
has shown that the attractor is periodic and the Lyapunov exponent $\lambda=0$.
Characteristic information on the dynamics gets encoded by periodic 
sequences $(\tau_m,\,L_{\rm act}^{s_m})$, $m=1,2,\dots$ with a period of some length $q$ 
such that $\tau_{m+q}=\tau_{m}$ and $s_{m+q}=s_{m}$ (for large enough $m$)
as in the example shown in Fig. (\figtimeseries) in the main article.  
If the dynamics of the system would remain in an active network $A_{\rm act}^s$ the Lyapunov exponent would be identical with the largest real part $L_{\rm act}^s$ of the eigenvalues of $A$.
The Lyapunov exponent $\lambda$ of the sequentially linear system therefore is well approximated\footnote{ 
	Convergence of $x_{\rm act}^s\to x^{*\,s}_{\rm act}$ or into the direction of the leading possibly complex eigenvector, 
	if $x^{*\,s}_{\rm act}$ is unstable, remains incomplete since convergence is always interrupted by a switching event.
} 
by the time average over $L_{\rm act}^{s_m}$, i.e.
\begin{equation}
\lambda\sim\lim_{m\to\infty} \frac{1}{Z_m}\sum_{n=1}^m \tau_n L_{\rm act}^{s_n}\,\quad Z_m=\sum_{n=1}^m \tau_n\,.
\label{eq:balance}
\end{equation} 
Since the dynamics is periodic the time average only needs to be taken over one period
and since $\lambda=0$ one gets $0\sim \sum_{k=1}^q \tau_{n+k} L_{\rm act}^{s_{n+k}}$ for $n$ large enough. 
The "refocusing" mechanism discussed above qualitatively therefore is also "balancing" the times $\tau_m$ specific active sets $s_m$ remain active 
by fine tuning switching times\footnote{ 
  This also is supported by the fact that simulations with finite time increment regularly
  produce chaotic dynamics with small but positive Lyapunov exponents since switching times can only 	
  be tuned to the accuracy of the time increment. However 
  $\lambda$ approaches zero consistently as the time increment is made smaller and orbits become
  periodic again. \vspace{0.2cm}
} such that contributions from time-domains with stable ($L_{\rm act}^{s_m}<0$) and 
unstable dynamics ($L_{\rm act}^{s_m}>0$) compensate each other. 
{\it Temporal balance} and {\it refocusing} 
are two aspects of the temporal self-organizing principle manipulating switching times.

\section*{Acknowledgments}
This work has been funded by the {\it Forum Integrativmedizin} an initiative of the {\it Hilde Umdasch Privatstiftung}.

\end{document}